 \documentclass[twocolumn,trackchanges]{aastex63}
\usepackage{CJK}
\usepackage{color}

\received{Month date, 2020}
\revised{Month date, 2020}
\accepted{Month date, 2020}
\shorttitle{Sample article}
\shortauthors{Hao Cheng et al.}

\begin{document}

\title{Measurements of $^{160}$Dy($p,\gamma$) at energies relevant for astrophysical $\gamma$ process}

\correspondingauthor{Bao-Hua Sun}
\email{bhsun@buaa.edu.cn}

\correspondingauthor{Li-Hua Zhu}
\email{zhulh@buaa.edu.cn}

\correspondingauthor{Yun Zheng}
\email{zhengyun@ciae.ac.cn}

\begin{CJK*}{UTF8}{gbsn}
\author{Hao Cheng}
\affiliation{School of Physics, and International Research Center for Big-Bang Cosmology and Element Genesis, Beihang University, Beijing 100191, P. R. China}
\affiliation{China Institute of Atomic Energy, Beijing 102413,  P. R. China\\}

\author{Bao-Hua Sun}
\affiliation{School of Physics, and International Research Center for Big-Bang Cosmology and Element Genesis, Beihang University, Beijing 100191, P. R. China}

\author{Li-Hua Zhu}
\affiliation{School of Physics, and International Research Center for Big-Bang Cosmology and Element Genesis, Beihang University, Beijing 100191, P. R. China}

\author{Motohiko Kusakabe}
\affiliation{School of Physics, and International Research Center for Big-Bang Cosmology and Element Genesis, Beihang University, Beijing 100191, P. R. China}
\author{Yun Zheng}
\affiliation{China Institute of Atomic Energy, Beijing 102413,  P. R. China\\}

\author{Liu-Chun He}
\affiliation{School of Physics, and International Research Center for Big-Bang Cosmology and Element Genesis, Beihang University, Beijing 100191, P. R. China}
\author{Toshitaka Kajino}
\affiliation{School of Physics, and International Research Center for Big-Bang Cosmology and Element Genesis, Beihang University, Beijing 100191, P. R. China}
\affiliation{National Astronomical Observatory of Japan, 2-21-1 Osawa, Mitaka, Tokyo 181-8588, Japan}
\affiliation{Graduate School of Science, The University of Tokyo, 7-3-1 Hongo, Bunkyo-ku, Tokyo 113-0033, Japan}

\author{Zhong-Ming Niu}
\affiliation{School of Physics and Material Science, Anhui University, Hefei 230039, P. R. China\\}

\author{Tian-Xiao Li}
\affiliation{China Institute of Atomic Energy, Beijing 102413,  P. R. China\\}

\author{Cong-Bo Li}
\affiliation{China Institute of Atomic Energy, Beijing 102413,  P. R. China\\}

\author{Dong-Xi Wang}
\affiliation{China Institute of Atomic Energy, Beijing 102413,  P. R. China\\}

\author{Meng Wang}
\affiliation{School of Physics, and International Research Center for Big-Bang Cosmology and Element Genesis, Beihang University, Beijing 100191, P. R. China}

\author{Guang-Shuai Li}
\affiliation{School of Physics, and International Research Center for Big-Bang Cosmology and Element Genesis, Beihang University, Beijing 100191, P. R. China}

\author{Kang Wang}
\affiliation{School of Physics, and International Research Center for Big-Bang Cosmology and Element Genesis, Beihang University, Beijing 100191, P. R. China}

\author{Lin Song}
\affiliation{School of Physics, and International Research Center for Big-Bang Cosmology and Element Genesis, Beihang University, Beijing 100191, P. R. China}

\author{Ge Guo}
\affiliation{School of Physics, and International Research Center for Big-Bang Cosmology and Element Genesis, Beihang University, Beijing 100191, P. R. China}

\author{Zhi-Yong Huang}
\affiliation{China Institute of Atomic Energy, Beijing 102413,  P. R. China\\}

\author{Xiu-Lin Wei}
\affiliation{School of Physics, and International Research Center for Big-Bang Cosmology and Element Genesis, Beihang University, Beijing 100191, P. R. China}

\author{Fu-WeI Zhao}
\affiliation{School of Physics, and International Research Center for Big-Bang Cosmology and Element Genesis, Beihang University, Beijing 100191, P. R. China}

\author{Xiao-Guang Wu}
\affiliation{China Institute of Atomic Energy, Beijing 102413,  P. R. China\\}

\author{Yimuran Abulikemu}
\affiliation{School of Physics, and International Research Center for Big-Bang Cosmology and Element Genesis, Beihang University, Beijing 100191, P. R. China}

\author{Jian-Cheng Liu}
\affiliation{China Institute of Atomic Energy, Beijing 102413,  P. R. China\\}

\author{Ping Fan}
\affiliation{China Institute of Atomic Energy, Beijing 102413,  P. R. China\\}





\begin{abstract}

Rare information on photodisintegration reactions of nuclei with mass numbers $A \approx 160$ at astrophysical conditions
impedes our understanding of the origin of $p$ nuclei.
Experimental determination of the key ($p,\gamma$) cross sections has been playing an important role in verifying nuclear reaction models and providing rates of  relevant ($\gamma, n$) reactions in $\gamma$ process.
In this paper we report the first cross-section measurements of $^{160}$Dy($p,\gamma$)$^{161}$Ho and $^{161}$Dy($p,n$)$^{161}$Ho in the beam energy range of 3.4 - 7.0 MeV, partially covering the Gamow window.
Such determinations are possible by using two targets with various isotopic fractions.
The cross section data can put a strong constraint on the nuclear level densities and gamma strength functions for $A \approx$ 160 in the Hauser-Feshbach statistical model.
Furthermore, we find the best parameters for TALYS that reproduce the  available A $\thicksim$ 160 data,  $^{160}$Dy($p,\gamma$)$^{161}$Ho and $^{162}$Er($p,\gamma$)$^{163}$Tm,
and recommend the constrained $^{161}$Ho($\gamma, n$)$^{160}$Dy reaction rates over a wide temperature range for $\gamma$ process network calculations.
Although the determined $^{161}$Ho($\gamma$, p) stellar reaction rates at the temperature of 1 to 2 GK can differ by
up to one order of magnitude from the NON-SMOKER predictions,
it has a minor effect on the yields of $^{160}$Dy and accordingly the $p$ nuclei, $^{156,158}$Dy.
A sensitivity study confirms that the cross section of $^{160}$Dy($p$, $\gamma$)$^{161}$Ho is measured precisely
enough to predict yields of $p$ nuclei in the $\gamma$ process.
\end{abstract}



\section{Introduction} \label{intro}
\end{CJK*}
The $\gamma$ process~\citep{Woosley1978The}, also referred to as $p$-process, was proposed as one of the most promising candidates
for producing more than 30 stable neutron-deficient isotopes in the region between $^{74}$Se and $^{196}$Hg.
It is essentially made of photodisintegrations of the ($\gamma,n$), ($\gamma, n$), or ($\gamma,\alpha$) types
by burning the preexisting seed $s$-isotopes in stellar environments
of high enough temperature  ($T_9 \equiv T /(10^9~{\rm K}) =[2,3]$) and subsequent $\beta^+$-decay and electron capture, possibly complemented with captures of neutrons, protons or
$\alpha$ particles at center-of-mass energies typically far below 1 MeV or the Coulomb barrier in the case of charged particles~\citep{Arnould2003The,2003Proton}.
Stable nuclei and nearby unstable nuclei are located on pathways on nuclear chart during the $\gamma$ process.
The required temperature conditions are fulfilled in the oxygen-/neon-rich layers of  type II supernovae \citep{Woosley1978The,ray90,pra90,ray95,cos00,2001Nucleosynthesis,2004PhRvL..93p1102H,iwa05,hay08} of massive stars or type Ia supernovae \citep{how91,how92,gor02,gor05,arn06,Kusakabe2010Production,2011ApJ...739...93T} of intermediate mass stars.

A reliable modeling of $\gamma$ process flows involves typically an extended network of about 2000 nuclei and 10,000 reactions for nuclei with masses up to around 210~\citep{Arnould2003The}.
The nucleosynthesis develops in the neutron-deficient region of the chart of nuclides~\citep{how91,ray95,Kusakabe2010Production}, whose masses and decay rates are mostly well known experimentally.
The reaction rates of such critical reactions, in particular for the photodissociations,
then represent the main uncertainty from the nuclear physics point view in predicting the elemental abundances produced in $\gamma$ process~\citep{2016Uncertainties,Nishimura2017Uncertainties}.

Despite decades of considerable experimental effort, the scarcity of the experimental reaction rates for $\gamma$ process
makes it mandatory to rely heavily on
the Hauser-Feshbach (HF) statistical model~\citep{Hauser1952The,Rauscher1997Nuclear}  and the various nuclear ingredients in such a framework.
In reality, it remains a challenge to validate the reliability of the HF model and its inputs, and accordingly to put the $\gamma$ process calculations on a more reliable base.
Therefore, providing crucial new experimental cross sections relevant to $\gamma$ process will play a key role (see, e. g.,~\citet{2003PhRvC..67a5807U,2006PhRvC..74b5806U,Dillmann2006,2006EPJA...27S.129D,2010PhRvC..81a5801D,2007PhRvC..76e5807K,2007EPJA...32..357M,2014PhRvC..90c5806N,Guray2015,2016PhRvC..93d5810L,2016PhRvC..93b5804H,2020PhLB..80535431W}.)
Experimentally, most of the photodissociations in the $\gamma$ process nucleosynthesis studies have been so far derived from the reverse radiative capture
on the basis of the reciprocity theorem.

Experiments in the $A \sim$ 160 mass region are particularly rare due to the extremely low cross section at the energies of Gamow window and the difficulty in target material purifications.
The $\gamma$ nuclides in this mass region are most efficiently produced in the stellar region which experiences peak temperature of  $T_9 =$ 2.6--2.85~\citep{Kusakabe2010Production}.
$^{162}$Er($p,\gamma$), the only data available in this range was studied recently using the activation method~\citep{2017as}.

Strong constraints for reaction models are helpful in improving and guaranteeing reliable results of the $\gamma$ process calculations.
The present paper describes the first measurements of $^{161}$Dy($p,n$)$^{161}$Ho and $^{160}$Dy($p,\gamma$)$^{161}$Ho, 
partially covering the Gamow window in the $\gamma$ process.
They are used to determine the ($\gamma$, $p$) rates to confirm predictability of existing models for reaction cross section and obtain realistic theoretical results of $\gamma$ process.
The measurements and the data analysis are described in Section 2.
In Section 3, we compare the data with TALYS~\citep{Koning2008TALYS} calculations obtained with different level density, gamma strength function, optical model, compound nucleus and pre-equilibrium. Moreover, we find the best calculation that can reproduce the experimental data available in the $A=160$ range, and derive the stellar photodisintegration rate for $^{160}$Dy of relevance for nucleosynthesis calculations, and a sensitivity calculation is made.
Final conclusions are drawn in Section 4.


\section{ Experiment and data analysis}\label{exp}

The experiment was performed with the activation method.
This method is suitable for radioactive reaction products with
half-life between several years and tens of minutes.
One counts the number of radioactive products via the activity after the irradiation took place.
This method is free of prompt beam-induced background at the irradiation, and avoids the details of decay branching or angular correlation effects of $\gamma$ transitions in the prompt $\gamma$ measurement.

\subsection{Target preparation}
Natural dysprosium is composed of seven stable isotopes with mass numbers 156, 158, 160, 161, 162, 163 and 164.
$^{156}$Dy and $^{158}$Dy have only tiny natural abundances and reactions on them are out of reach in the current experiment conditions.
Of all the ($p,\gamma$) products, $^{165}$Ho is stable and $^{163}$Ho has a half-life of 4570(20) years thus is too long to be studied.
At 1.65 MeV proton bombarding energy the $^{161}$Dy($p,n$)$^{161}$Ho reaction channel opens, which results in the same final nucleus as $^{160}$Dy($p,\gamma$)$^{161}$Ho.
The threshold energies for the $^{162}$Dy($p,n$)$^{162}$Ho, $^{163}$Dy($p,n$)$^{163}$Ho, and $^{164}$Dy($p,n$)$^{164}$Ho are 2.94, 0.79, and 1.78 MeV, respectively.
Hence using natural or enriched but not 100\% abundance isotope Dy targets above the ($p,n$) reaction thresholds,  ($p,n$) and ($p,\gamma$) reaction channels cannot be distinguished, and the resulting cross section is the weighted sum of the two cross sections.

\begin{table*}[htpb!]
\centering\caption{Isotopic abundances in the natural Dy ($^\textrm{nat}$Dy) and the enriched $^{160}$Dy targets ($^{160}$Dy$_2$O$_3$).} \label{tab:1}
\begin{tabular}{cccccc}
\hline\hline
Type & Material & Isotopic Composition  \\
\hline \rule{0em}{10pt}

1    & $^\textrm{nat}$Dy       &  $^{160}$Dy(2.34\%), $^{161}$Dy(18.91\%), $^{162}$Dy(25.51\%), $^{163}$Dy(24.90\%), $^{164}$Dy(28.18\%) &                  \\
2    & $^{160}$Dy$_2$O$_3$       &  $^{160}$Dy(51.82\%), $^{161}$Dy(13.87\%), $^{162}$Dy(5.79\%), $^{163}$Dy(3.05\%), $^{164}$Dy(1.68\%) &                 \\

\hline \hline

\end{tabular}
\end{table*}

One way to extract the ($p,\gamma$) cross sections is to employ two kinds of targets with different isotopic abundances.
By coupling the ($p,\gamma$) and ($p,n$) channels to the same radioactive nuclide, the individual cross section can be deduced in the case of good statistics.
The information of two targets used is summarized in Table~\ref{tab:1}, one with natural Dy and the other with $^{160}$Dy-enriched abundance.
With the abundance distributions in two targets, we can determine the cross sections for the two reactions $^{160}$Dy($p,\gamma$)$^{161}$Ho and $^{161}$Dy($p,n$)$^{161}$Ho.

The thicknesses of natural targets are around 1-2 mg cm$^{-2}$ and those of the enriched targets are 2 mg/cm$^{-2}$ as displayed in Table~\ref{measurement}.
These targets are ellipses with an axis length of 9 mm $\times$ 12 mm, and the sizes of target frames are 25 mm $\times$ 12.5 mm.
The self-supported natural dysprosium targets were made by rolling method. Enriched $^{160}$Dy (51.82\%) targets were sputtered onto a $^{197}$Au backing.
The mass abundances of contamination elements are determined typically to be less than 0.1\% and 1\% for the natural targets and enriched targets by the energy-dispersive X-ray spectroscopy, respectively.

\subsection{Experiment Setup}
The investigated energies in this work ranging from 3.4 to 7 MeV are above the relevant ($p,n$) reaction thresholds.
These energies partially cover the Gamow window for the $\gamma$ process in this mass region for $T_9$ = [2, 2.9].
\begin{center}
\includegraphics[width=0.48\textwidth,clip=true,trim=0cm 0cm 0cm 0cm]{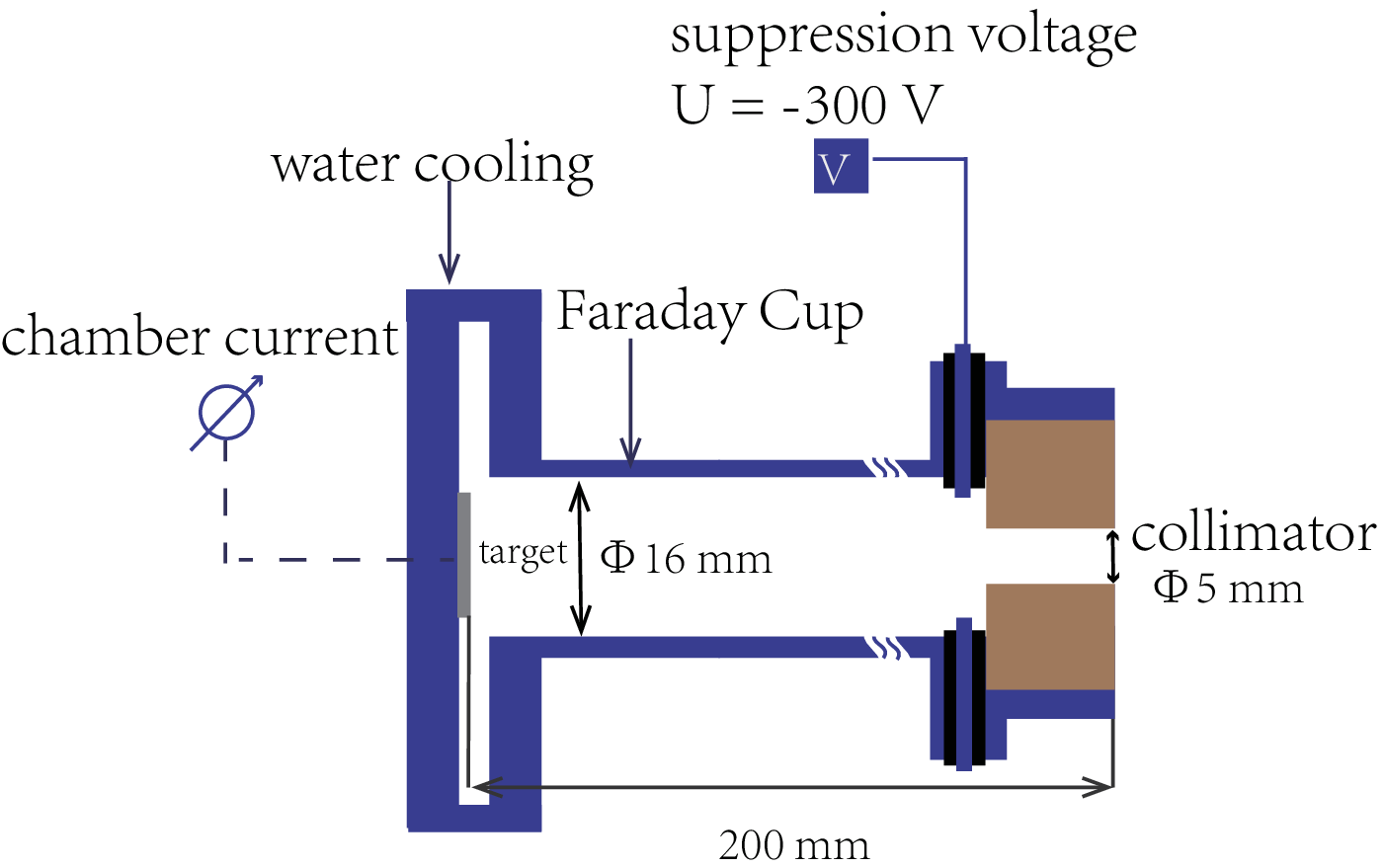}
\figcaption{\label{fig1} Schematic drawing of the target chamber. Refer to the text for details.}
\end{center}

The proton beams were delivered by
2$\times$1.7 MeV and H1-13 tandem accelerator at China Institute of Atomic Energy, Beijing, respectively.
The maximum energy provided by the 2$\times$1.7 MeV tandem accelerator and the minimum energy that
can be steadily and intensely provided by the H1-13 tandem accelerator are 3.4 MeV and 5 MeV, respectively.
The proton current throughout the irradiations was between 150 nA and 700 nA for different beam energies.

A schematic drawing of the target chamber is shown in Figure \ref{fig1}. The target chamber also servers as a Faraday cup with a $\Phi$5 mm-hole collimator and insulated from the beam tube. The distance from the hole collimator to the target surface is 20 cm. A suppression voltage of -300 volts was applied to suppress secondary electrons
emitted from the target. 
A water-cooling system for the target chamber was used to maintain the temperature and minimize surface deformation of the target. The beam current was digitized with a frequency of 1 Hz by the Keithley 6517B electrometer.

Figure \ref{fig2} shows the beam current during the beam time.
The variance in beam currents $I(t)$ can bring an up to 35\% difference in the cross section when considering the maximum and minimum beam intensity. This indicates the necessity of recording the precision current with time.
\begin{center}
\includegraphics[width=0.52\textwidth,clip=true,trim=0cm 0cm 0cm 0cm]{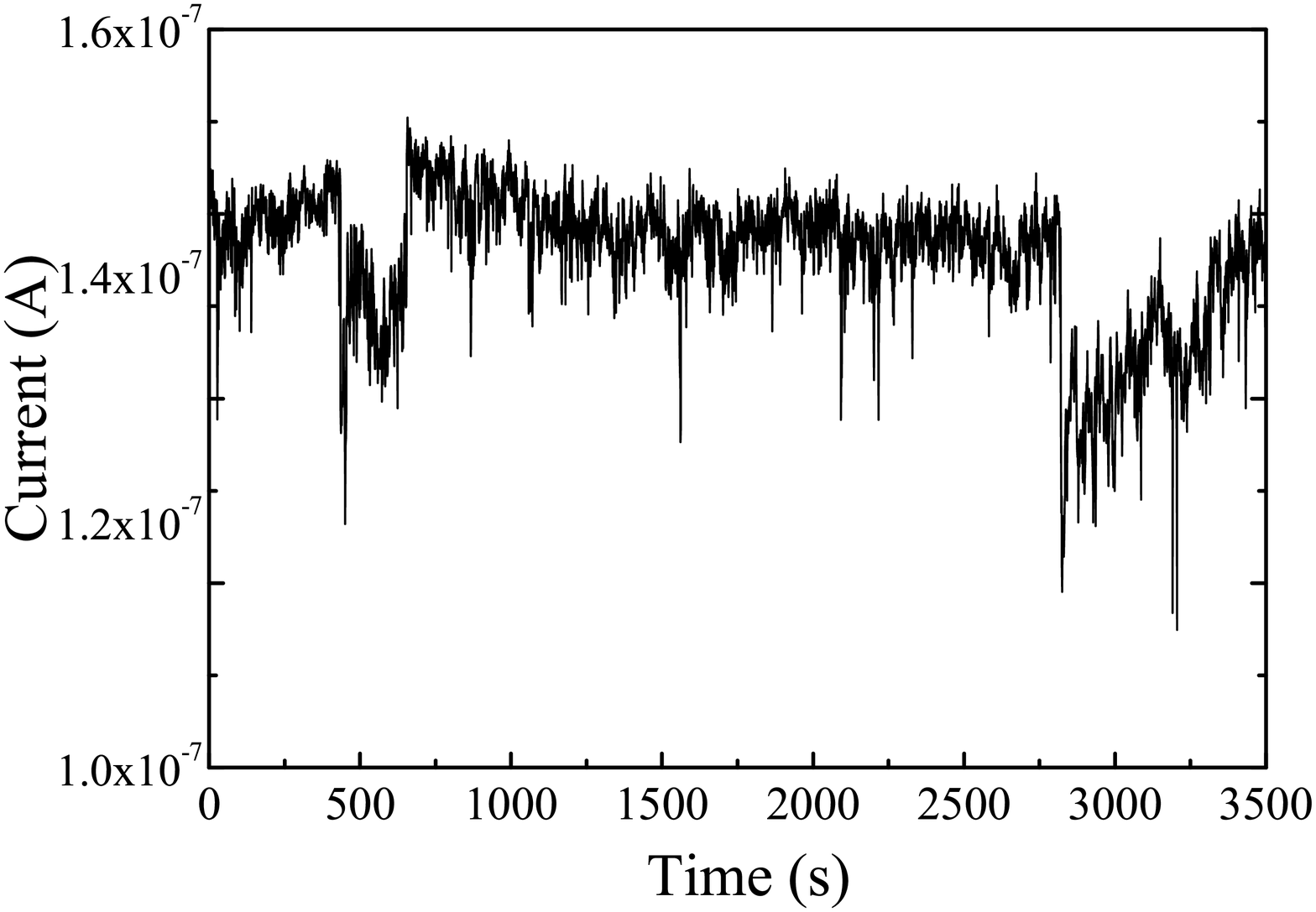}
\figcaption{\label{fig2}Variation of the beam current during the beam time.}
\end{center}

The effective center-of-mass energy ($E_{\text{c.m.}}$) for each run is calculated by considering the proton energy loss  $\Delta E$ in the target material and follows
\begin{equation}
E_{\text{c.m.}}= \frac{M_{A}}{M_{p}+M_A}(E_{p}-\frac{\Delta E}{2}) \; ,
\end{equation}
where $M_{A}$ and $M_{p}$ are the mass of the target nuclide and the proton, $E_{p}$ is the proton energy.
Energy losses are calculated using ATIMA, a built-in Physical Calculator in LISE++~\citep{Tarasov2008LISE}.
The energies used in the experiments are summarized in Table~\ref{measurement}. The relevant uncertainties include the energy straggling and the difference caused by using different thickness of targets.

The targets were irradiated for 30-60 minutes (irradiation time) and then followed by 30-1000 minute measurements (measurement time), while in between there were 10 to 20 minutes (waiting time) to release the vacuum, dismount the target and place the target in the position for off-line measurement. The details of measurement are displayed in Table~\ref{measurement}.

\begin{table*}[htpb!]
\centering\caption{Details of the Measurements. Target and it thickness (in units of mg cm$^{-2}$), and the center-of-mass energy $E_\textrm{c.m.}$ (in units of MeV), the relevant irradiation time (in units of seconds), waiting time  (in units of seconds), measurement time  (in units of seconds) are listed. }\label{measurement}
\begin{tabular}{cccccc}
\hline\hline
Target  & Thickness & $E_\textrm{c.m.}$& Irradiation Time &Waiting Time&  Measurement Time\\
     &    (mg cm$^{-2}$)   &(MeV)    &(s) & (s) & (s) \\
\hline \rule{0em}{10pt}
$^\textrm{nat}$Dy     &  1.92(10)   & 3.34(8)       & 2484 & 620& 3311  \\
$^{160}$Dy$_2$O$_3$  & 1.85(19)     & 3.33(9)   &2157&796&55297\\
$^\textrm{nat}$Dy     &  1.92(10)   & 4.94(6)     & 3524 & 745& 4742  \\
$^{160}$Dy$_2$O$_3$   & 2.13(21)    & 4.92(8)  &3767&814&9280\\
$^\textrm{nat}$Dy     &  1.72(9)     & 5.44(5)    & 3551 & 1046& 4725  \\
$^{160}$Dy$_2$O$_3$   & 1.92(19)      & 5.42(7)   &3597&1172&9555\\
$^\textrm{nat}$Dy     &  0.96(5)      & 5.95(3)   & 2108 & 1084& 2776  \\
$^{160}$Dy$_2$O$_3$    & 2.13(21)     & 5.92(7) &2059&1011&7973\\
$^\textrm{nat}$Dy      &  1.07(5)     & 6.45(3)   & 1757 & 1175& 3307  \\
$^{160}$Dy$_2$O$_3$     & 1.60(16)    & 6.43(7)  &2059&934&7714\\
$^\textrm{nat}$Dy      &  0.92(5)     & 6.95(2)      & 1182 & 840& 2608  \\
$^{160}$Dy$_2$O$_3$    & 1.85(19)     & 6.92(5)  &1528&832&2165\\ 
\hline \hline

\end{tabular}
\end{table*}

The $\gamma$ radiations following the electron-capture decays of the produced Ho isotopes were measured with a CLOVER detector 
in a low background shielding system as shown in Figure~\ref{fig3}.
The CLOVER consists of four coaxial $N$-type high-purity Germanium detectors, each with a diameter of 60 mm and a length of 60 mm. The energy resolution for the CLOVER is 2.2 keV (FHWM) for the 1.332 MeV $\gamma$ rays of $^{60}$Co.
The relative efficiency is 38\% for each Germanium crystal.
The shielding system is a cylinder with a radius of 64 cm and a height of 66.1 cm. It consists of layers of iron, lead, copper, and plexiglass from the outside to the inside.
The lead layer is used to shield most of the low energy environment background, and the copper layer aims to absorb the characteristic X-rays of lead.
This system has been used to investigate the intrinsic radiation background of a B380 LaBr$_3$(Ce) detector~\citep{Cheng2020Intrinsic}.

\begin{center}
\includegraphics[width=0.30\textwidth,clip=true,trim=0cm 0cm 0cm 0cm]{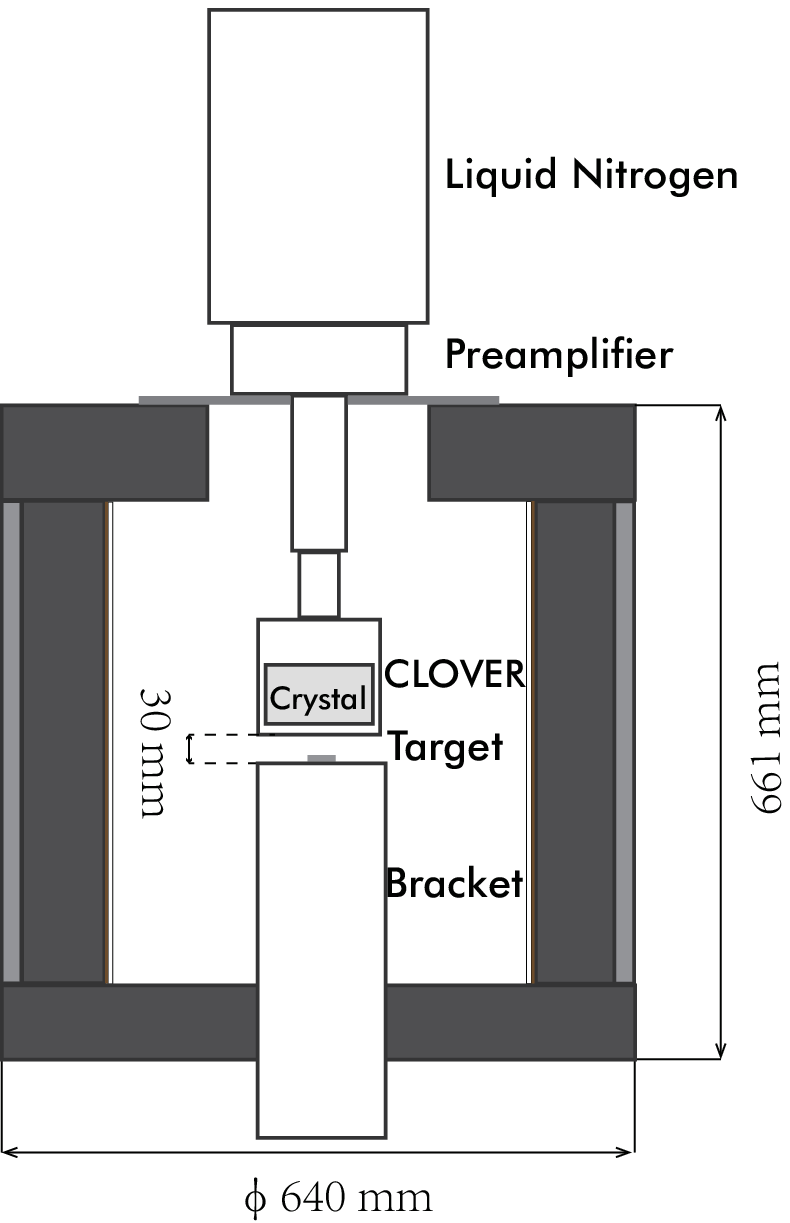}
\figcaption{\label{fig3} Schematic drawing of the $\gamma$ activity setup. It contains a CLOVER crystal and a low-background shielding room. The shielding room is composed of plexiglass, copper, lead, and iron from the inside to the outside. Their relevant layer thicknesses are 5 mm, 2 mm, 84 mm and 20 mm, respectively. The irradiated targets were placed 3 cm away from the front surface of the detector.}
\end{center}

The irradiated targets were placed 3 cm away from the front surface of the detector.
The absolute efficiency was calibrated with one $^{152}$Eu and six  monoenergetic sources ($^{137}$Cs, $^{241}$Am, $^{54}$Mn, $^{88}$Y, $^{104}$Cd, and $^{65}$Zn).
The sources were placed at the same position as the targets.
The absolute full-energy peak efficiency curve was described using the EFFIT program in the Radware \citep{Radford1995ESCL8R} package.
The efficiency curve was further corrected for the summing coincidence effect by a dedicated GEANT4 simulation~\citep{He2018Summing}.
The impact of geometric acceptance due to the size of irradiation target, is estimated to be less than 1\%. This effect is much smaller than the statistic error.

Signals of the CLOVER detector were recorded by the VME data acquisition system (DAQ). The timestamp of each event was digitized in a precision of 10 ms. This information
is useful to determine the decay curve of a specified $\gamma$ activity and can be used as an independent check of the radioactive product.
Dead time correction was also added in DAQ for the absolute counting.

\subsection{Cross-section Determination}

There is a low-lying  1/2$^+$ isomeric state with a half-life of 6.67 (7) s in $^{161}$Ho. Nevertheless, this
isomer does not affect the determination of cross sections because its half-life is much shorter than the irradiation time of typically more than half an hour. All the isomeric states would have decayed to the ground state by isomeric transition (IT) in the off-line activity measurement.

\begin{figure*}
\begin{center}
\includegraphics[width=1.10\textwidth]{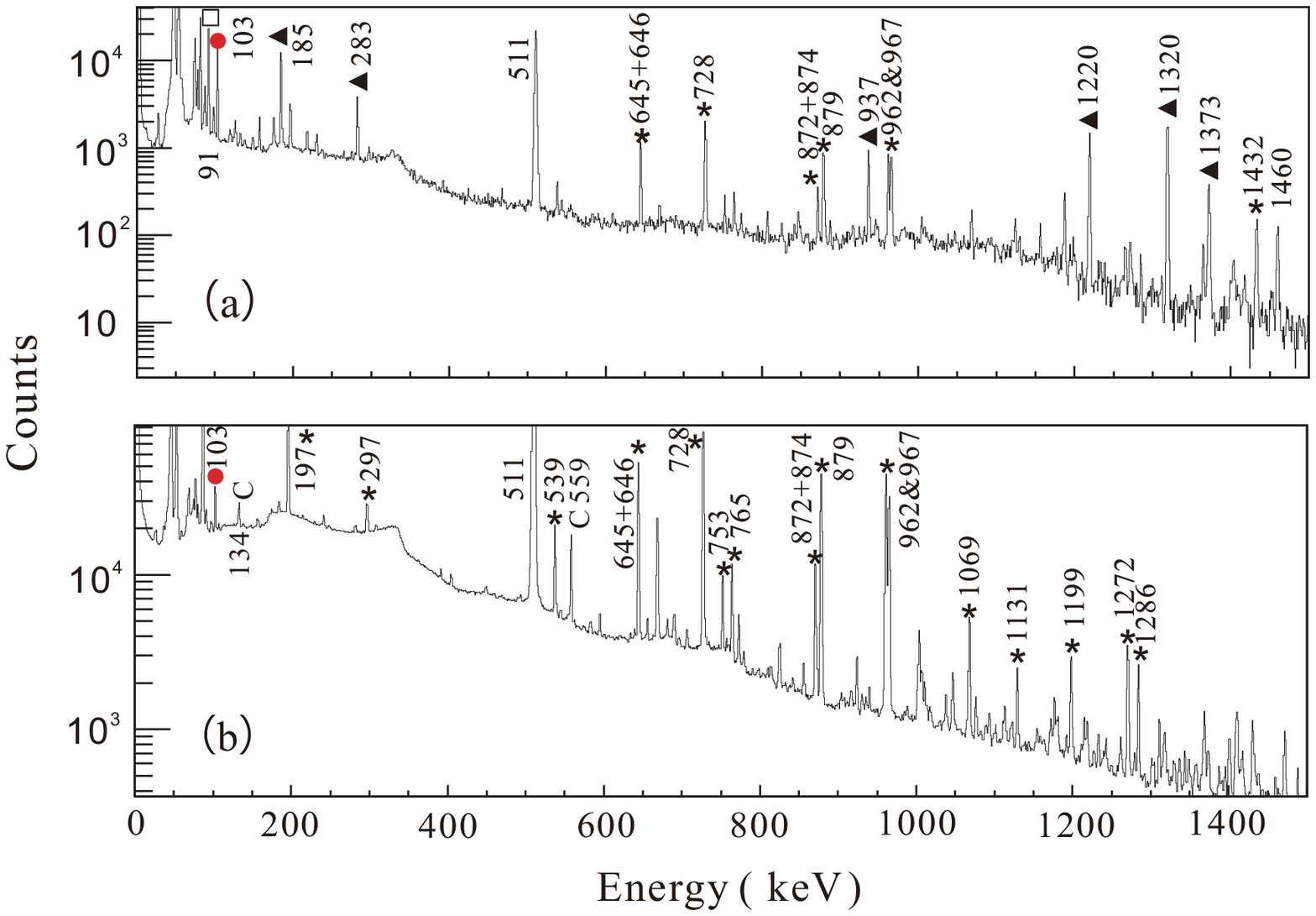}
\end{center}
\caption{Activation $\gamma$-ray spectra below 1500 keV
taken from the measurements using the natural Dy target (a) and the enriched $^{160}$Dy target (b)
at the proton beam energy of 6.5 MeV. 
Labeled are the characteristic $\gamma$-rays from $^{160}$Ho (asterisk), $^{161}$Ho (circles), $^{162}$Ho (triangles), and $^{164}$Ho (box) , and the energies marked by C are contaminants. The X-rays from Ho, Dy, and Au at the low energy are also visible. 
\label{fig4}}
\end{figure*}


Figure \ref{fig4} shows the typical activation $\gamma$-ray spectra below 1500 keV taken from the natural Dy and enriched Dy targets irradiated by 6.5 MeV protons.
We identified the characteristic $\gamma$-rays at 103 keV from $^{161}$Ho, the $\gamma$-rays at 197, 728, 879, etc., from $^{160}$Ho, the $\gamma$-rays at 185, 283, 937, 1220, 1320, and 1373 keV from $^{162}$Ho,
and the $\gamma$-ray at 91 keV from $^{164}$Ho.
The 103-keV $\gamma$ transition with a
relative intensity of 3.9\% per decay of $^{161}$Ho was used for data analysis.
The purity of the $\gamma$ ray is verified by examining its time evolution. The deduced half-life of 148.5(24) minutes is in a good agreement with 148.8(30) minutes reported in Ref.~\citep{2011Nuclear}

The count of the 103 keV $\gamma$ peak was analyzed using the Radware package \citep{Radford1995ESCL8R} and corrected with the dead time of DAQ.
Decay parameters of Ho isotopes, the decay constant ($\lambda$) and the relative intensity ($\eta_\gamma$) of the characteristic $\gamma$ ray, are from the Nuclear Data Sheets (NDS; ~\citep{Singh2014Nuclear}).
By decoupling the total counts of characteristic $\gamma$ rays measured in both types of targets at the same proton energies, we can deduce the cross sections for $^{160}$Dy($p,\gamma$)$^{161}$Ho and $^{161}$Dy($p,n$)$^{161}$Ho. The derivation of cross section from the activation $\gamma$ rays is described in detail in Section appendix .

\section{Results and discussion}

\begin{table}
\footnotesize
\centering\caption{Determined cross sections and astrophysical $S$ factors at the relevant Center-of-mass energy $E_{\text{c.m.}}$. }\label{tab:2}
\begin{tabular}{ccccc}
\hline\hline
 $E_{c.m.}$& $\sigma_{^{160}\textrm{Dy}(p,\gamma)^{161}\textrm{Ho}}$ & $S$ Factor &$\sigma_{^{161}\textrm{Dy}(p,n)^{161}\textrm{Ho}}$ \\
  (MeV)    &(mb)                   & (MeV b)             &(mb)\\
\hline \rule{0em}{10pt}
    3.34(6)   & 3.25(58)$\times10^{-3}$  & 3.22$\times10^{10}$ & 3.72(70)$\times10^{-3}$ \\
   4.93(5)    &  0.106(18)                & 2.84$\times10^{9}$  &    0.60(11) \\
   5.43(4)    & 0.152(25)                 & 1.09$\times10^{9}$  &    1.60(30)   \\
   5.94(4)    &  0.243(40)                & 5.72$\times10^{8}$  &    4.47(84)    \\
   6.44(3)    &  0.310(51)               & 2.79$\times10^{8}$  &    9.15(17)\\
   6.93(3)    &  0.507(84)               & 1.92$\times10^{8}$  &    19.4(36)  \\
\hline \hline
\end{tabular}
\end{table}

The cross sections of $^{160}$Dy($p,\gamma$)$^{161}$Ho and $^{161}$Dy($p,n$)$^{161}$Ho are summarized in Table \ref{tab:2}.
The uncertainties in cross sections come from the propagation of errors in $\gamma$-ray counts (0.6\%$\thicksim$6.4\%),
gamma intensity from NDS (15\%), half-life of characteristic $\gamma$-rays (2\%), target thicknesses (5\% for the natural target and 10\% for the enriched target),
and $\gamma$ detection efficiencies (3\%).
The total uncertainties are dominated by
the relative poor precision in $\gamma$ intensity.
The cross sections caused by the difference in the thicknesses of the natural and enriched targets is less than 5\%. This effect is taken into account by introducing the uncertainties of proton energies.
Moreover, as an independent examination of our experimental setup and data analysis, we performed a cross-section measurement of $^{63}$Cu($p,n$) at a proton energy of 9.2 MeV.
The target consisted of a 1.1 mg/cm$^{-2}$ foil of natural Cu with a backing of 6.14 mg/cm$^{-2}$ thick $^{197}$Au.
The deduced cross section of 0.38 (3) mb is consistent with 0.367 (27) mb at 9.2 MeV reported by \citet{1974Excitation}.

The experimental cross sections are compared to theoretical calculations using the codes
NON-SMOKER~\citep{Rauscher2001Tables}, TALYS ~\citep{Koning2008TALYS} and EMPIRE~\citep{Herman2007EMPIRE} in Figure ~\ref{fig5}.
The NON-SMOKER calculations are widely adapted in current network calculations and have been included in the JINA REACLIB database~\citep{Cyburt2010}. The TALYS and EMPIRE calculations were performed using the default parameters.
As for the ($p,\gamma$) reaction, all three calculations can reproduce the 3.34 MeV data and agree with each other nicely at the energy up to around 4 MeV, the neutron threshold, which corresponds to the excited state near the neutron separation energy of $^{161}$Ho. However, deviations present above the neutron threshold, particularly for the NON-SMOKER results, which overpredict the cross sections at 5 $\thicksim$ 7 MeV by a factor of 3-6. Moreover, NON-SMOKER predicts a kink point at around 5.5 MeV, which seems to be different from the experimental data.
The EMPIRE calculations reproduce nicely the ($p,n$) data and the 3.34 MeV ($p,\gamma$) data, but underestimate the experimental centroid values
by a factor of 2 at 5 $\thicksim$ 7 MeV although they are generally consistent within the experimental uncertainties.
Overall, a good agreement between the data and the TALYS calculations with default input parameters is observed for both reactions at the energies investigated.  However, we note that the calculated cross section of $^{162}$Er$(p,\gamma)^{163}$Tm using the same default parameters is two times smaller than the experimental data at 6.5 to 9 MeV~\citep{2017as}. 

\begin{center}
\includegraphics[width=0.52\textwidth,clip=true,trim=0cm 0cm 0cm 0cm]{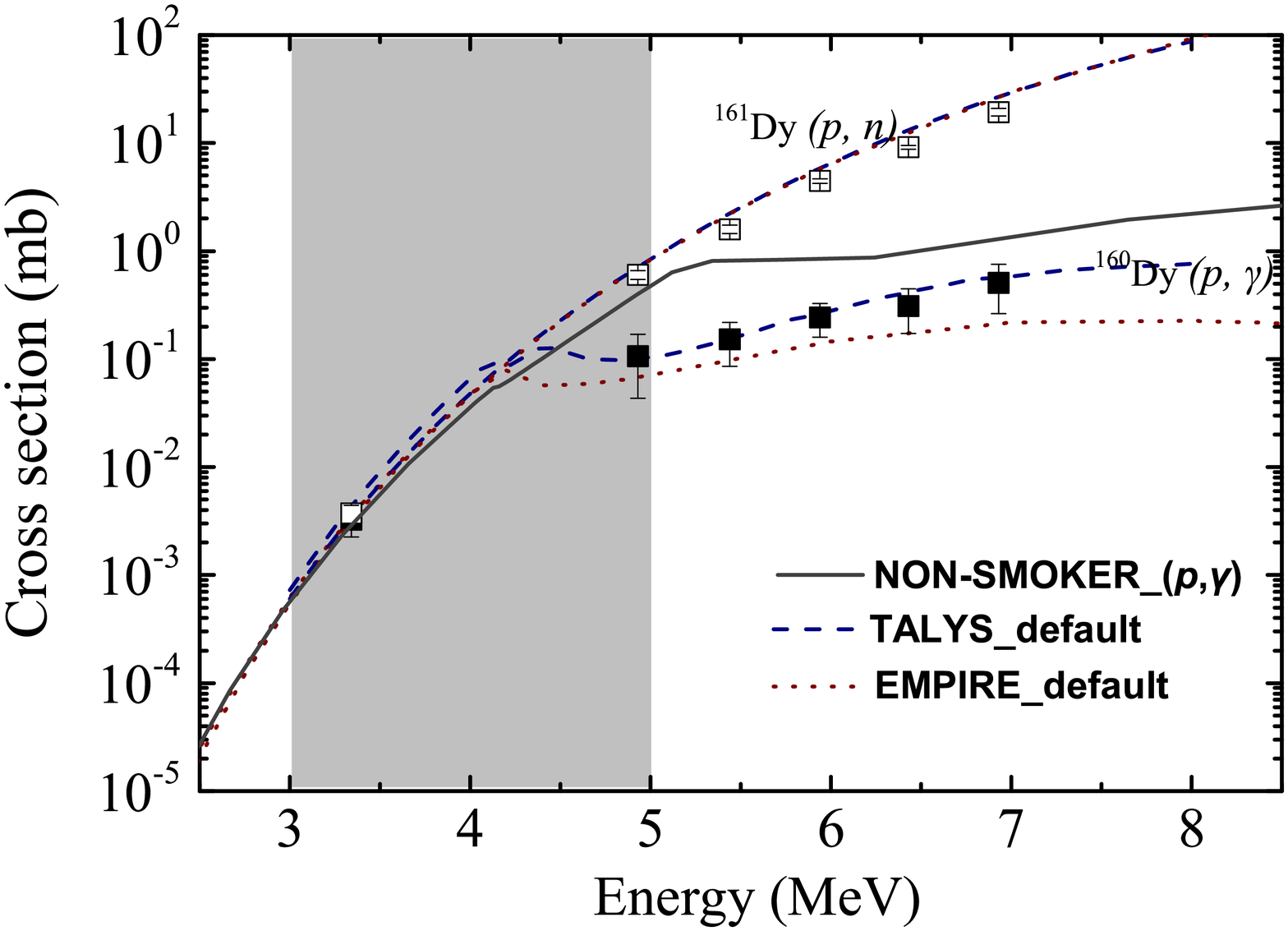}
\figcaption{\label{fig5} Measured cross sections of the $^{160}$Dy($p,\gamma$)$^{161}$Ho (solid square) and $^{161}$Dy($p,n$)$^{161}$Ho (open square) reactions compared with the predictions by the NON-SMOKER (solid line), EMPIRE (dotted line) code, and the TALYS (dashed line) codes using their default parameters. The relevant Gamow window for ($p,\gamma$) is indicated by the shadowed area. The threshold for $^{161}$Dy($p,n$)$^{161}$Ho is 1.65 MeV.}
\end{center}

\subsection{Constraining the Hauser-Feshbach model}\label{Modelcalculations}

The key ingredients in the Hauser-Feshbach calculations~\citep{Rauscher2013Constraining}
include $\gamma$ strength functions (GSF), optical model potential (OMP),  the nuclear level densities (NLD) and mass models.
TALYS offers a variety of options for the description of these inputs and is used hereafter to make the sensitivity studies~\citep{MeiBo2015,Sensitivity2006}.

In Figure \ref{fig6}(a), the ($p,\gamma$) experimental results are compared with the predictions by the TALYS-1.9 code utilizing different NLD models, 
while all the other inputs are kept to be the same as the default ones.
Using different NLDs 
gives similar trends, so we present the selective calculations including the one with the default parameters as well as those of the lowest and highest cross sections, as limits of these calculations.
The default NLD in TALYS is the Constant temperature $+$ Fermi gas (CTFG) level density~\citep{A1965A}. It sets the
lower limit of cross sections, while the upper limit is obtained by using the microscopic NLD from Hilaire's table (HT)~\citep{2008Drip},
As shown in Figure \ref{fig6}(a), the ($p,\gamma$) cross section is sensitive to the NLDs from about 4 MeV on, and has a less significant impact in the Gamow window, i.e., the low energy data.
Variation of the level density modes could result in a factor of 4.5 difference in cross section at 7 MeV.
Our data are close to the predicted lowest limit, and thus provide an direct constraint for the NLD and rule out many NLDs in this mass range.

\begin{center}
\includegraphics[width=0.52\textwidth,clip=true,trim=0cm 0cm 0cm 0cm]{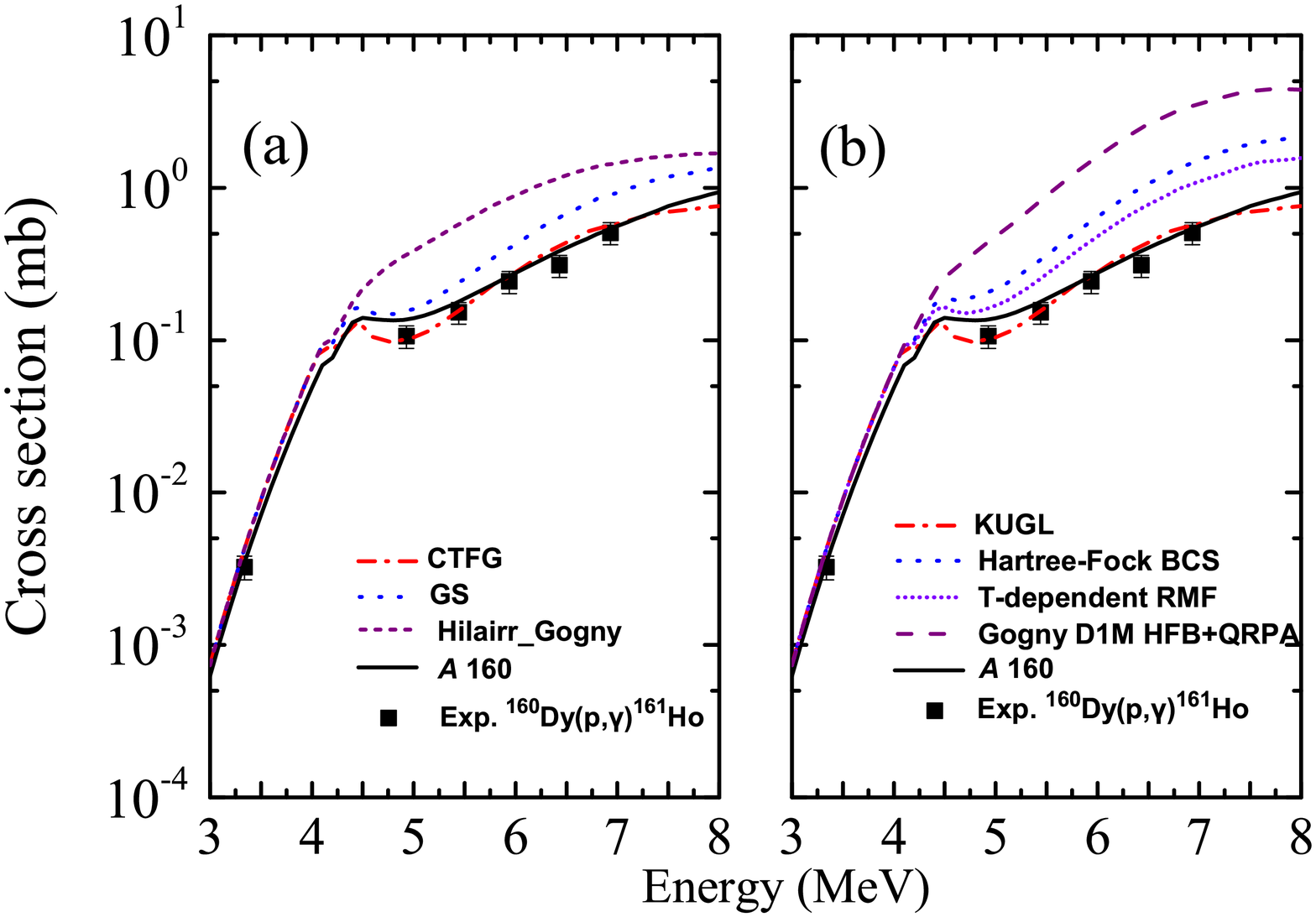}
\figcaption{\label{fig6} (a) Comparison of the measured $^{160}$Dy($p,\gamma$)$^{161}$Ho cross sections (squares) with the predictions of TALYS using different level density models
while keeping the other inputs as the default ones.  The solid line represents the optimized calculations  for $A \thicksim$ 160.  (b) Same as (a) but using different $\gamma$ strength functions. Refer to the text for details. }
\end{center}

Similarly, we present TALYS predictions using three GSFs in Figure ~\ref{fig6}(b).
The default GSF in the TALYS is the traditional standard Lorentzian model of Kopecky-Uhl generalized Lorentzian (KUGL)~\citep{1990Test}.
The lowest and highest cross sections are obtained by using the KUGL and the Gogny D1M HFB+QRPA models~\citep{Goriely2018The}, respectively.
Generally, the ($p$,$\gamma$) cross section depends sensitively on both GSF and NLD modeling.

\begin{table*}[htbp!]
\footnotesize
\centering\caption{Deduced Stellar Rate (s$^{-1}$) of the $^{161}$Ho($\gamma$, $p$)$^{160}$Dy in Temperatures $T_9$ = [0.1,10]. }\label{tab:4}
\begin{tabular}{cccc}
\hline\hline
$T_9$ &  Rate & $T_9$ &  Rate\\
\hline \rule{0em}{10pt}
0.10  & 6.89$\times10^{-279}$ &   1.50 &  3.41$\times10^{-12}$  \\
0.15 &  8.16$\times10^{-279}$ &   2.00 &  7.45$\times10^{-06}$ \\
0.2    & 3.17$\times10^{-279}$ &   2.50 &  6.78$\times10^{-02}$  \\
0.3    & 1.09$\times10^{-95 }$ &   3.00 &  3.22$\times10^{+01}$ \\
0.4  & 2.09$\times10^{-71 }$ &   4.00 & 6.46$\times10^{+04}$ \\
0.5  & 2.59$\times10^{-56 }$ &   5.00 & 5.18$\times10^{+06}$ \\
0.6  & 6.05$\times10^{-46 }$ &   6.00 & 9.81$\times10^{+07}$ \\
0.7  & 2.43$\times10^{-38 }$ &   7.00 & 9.71$\times10^{+08}$ \\
0.8  & 1.68$\times10^{-32 }$ &   8.00 & 6.07$\times10^{+09}$  \\
0.9   & 7.38$\times10^{-28 }$ &   9.00 &  2.41$\times10^{+10}$  \\
1.00 & 4.54$\times10^{-24 }$ &   10.00& 6.39$\times10^{+10}$  \\

\hline \hline
\end{tabular}
\end{table*}

We also examined the ($p, \gamma$) cross sections by using various compound models, pre-equilibrium models and optical models in TALYS. They have only a minor effect on the computed cross sections.
The optical potential is expected to be important at the astrophysically relevant low energies for charged particle captures and photodisintegrations. The one and only data point at 3.34 MeV can already help to exclude several models.

Further, it is worth noting that 
nuclear reaction calculations involve several (sub)models to compute the key input quantities like level density,  $\gamma$ strength function, optical potential, nuclear masses,  pre-equilibrium, and the compound process.
Their interplay makes the model evaluation almost impossible.
Instead of developing a consistent reaction model, we attempt to find the best model parameters of TALYS, which can well describe the
known experimental data. We are aware that the parameters might be mass range dependent or only hold to a certain precision in applying to nuclei nearby; therefore we concentrate on the mass range of 160.
This strategy is practically useful to reduce the nuclear physics
inputs in  $\gamma$ process simulations \citep{2016Uncertainties,Nishimura2017Uncertainties}.
We thus made a Monte Carlo calculation by randomly varying the key physical input parameters~\citep{2019TENDL},  
and employed the least-square method to find the calculation that matches best both the $^{160}$Dy($p,\gamma$)$^{161}$Ho and $^{162}$Er($p,\gamma$)$^{163}$Tm data. The results are shown in Figure~\ref{fig6}.


\subsection{Evaluation of the Stellar Rate}
 
The TALYS results constrained by the present experimental data, as presented in the previous section, represent a solid basis for the reaction rate calculation.
We thus employ the optimized parameters to compute the astrophysical rate of  $^{161}$Ho($\gamma, n$). It is obtained by integrating the cross section over a Maxwell-Boltzmann distribution of energies $E$ at the given temperature $T$.
In addition, the thermal excitation effect in hot astrophysical plasma is taken into account.

The calculated stellar reaction rates sharply depend on the temperature and
span over 250 order of magnitudes in the temperature range of $T_9$ = [0.1,1.0].
The rates are summarized in Table ~\ref{tab:4}.
Figure~\ref{fig7} shows
and the ratios to the NON-SMOKER predictions in the temperature of $T_9$ = [0.1,10].  The theoretical uncertainty~\citep{2019TENDL} is shown by the shadowed band. The ratio is up to one order of magnitude, and NON-SMOKER underproduced the rates at $T_9$ = [1, 2]. The difference would affect the yield of $^{160}$Dy in the $\gamma$ process and will be discussed in the next subsection.

\begin{center}
\includegraphics[width=0.52\textwidth,clip=true,trim=0cm 0cm 0cm 0cm]{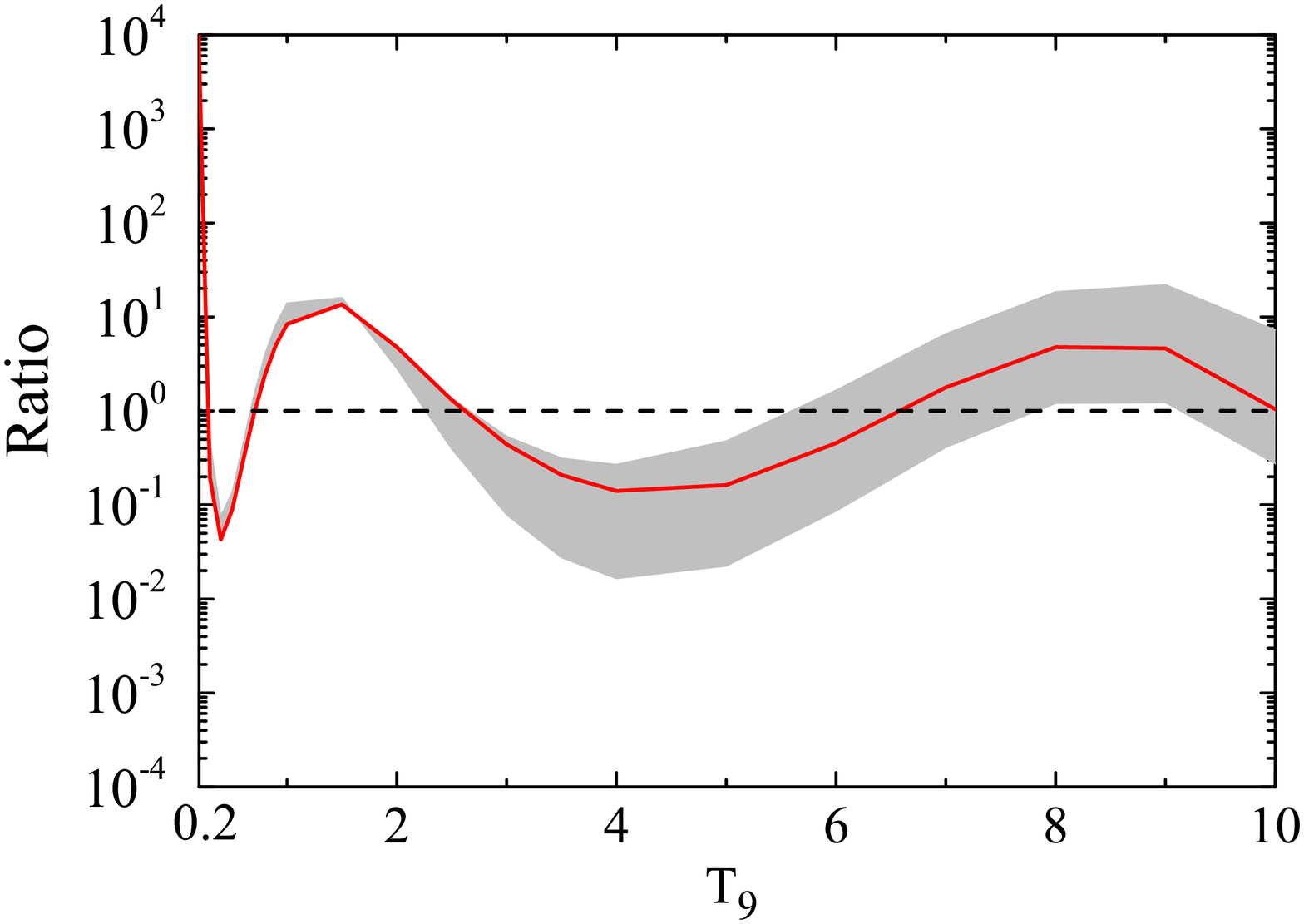}
\figcaption{\label{fig7}
The ratio of the deduced reaction rates (red solid line) to the NON-SMOKER predictions in the temperature of $T_9 = [0.2,10]$ . The shadowed area represents the theoretical uncertainty.}   
\end{center}

\subsection{Nucleosynthesis Calculation for $A \sim$ 160 $p$ nuclei}

As a possibly important region for the  $\gamma$ process of $A \sim 160$ region, we tested trajectories with peak temperatures $T_{9p} =[2,2.9]$ and peak density $\rho_{\rm p} =10^6$ g cm$^{-3}$ assuming the $\gamma$ process layer in Supernovae (SNe) Ia. They are assumed to decrease exponentially as $T_9 =T_{9p} \exp[-t/(3 \tau)]$ and $\rho =\rho_{\rm p} \exp(-t/\tau)$ with $\tau=1$ s. The initial nuclear composition is adopted from Case A1 of   \citet{Kusakabe2010Production} which is based on an efficient $s$-process during thermal pulses in the presupernova stage. The nucleosynthesis calculation is then performed with a variable-order network code \citep{Kusakabe2019Supernova} which is updated with new theoretical estimates on the rates of $^{16}$O($n$,$\gamma$)$^{17}$O \citep{2020Nuclear} and $^{17}$O($n$,$\gamma$)$^{18}$O \citep{Zhang2021}.

In the SNe $\gamma$ process environment, a sudden heating followed by a nearly adiabatic cooling occurs. The heating is triggered either by a deflagration or detonation wave in thermonuclear explosion corresponding to SNe Ia, or a shock in gravitational collapse type explosion corresponding to SNe II. Then, early in the heating epoch, first neutron source nuclei are burned to proceed with the neutron-capture process. As a result, neutron-rich nuclei are produced from seed $s$ nuclei. As the temperature increases further, the ($\gamma$, $n$) reactions start operating, and abundant nuclei shift from neutron-rich nuclei to neutron-deficient nuclei. If the peak temperature is high enough, ($\gamma$, $p$) and ($\gamma$, $\alpha$) reactions also work and heavy nuclei are disintegrated to lighter nuclei \citep[e.g.,][]{Kusakabe2010Production}. We estimate a sensitivity of nuclear yields to the reaction rate of $^{161}$Ho($\gamma, n$)$^{160}$Dy. Any effect of changing the $^{161}$Ho($\gamma, n$)$^{160}$Dy rate always occurs via changes of $^{161}$Ho and/or $^{160}$Dy abundances in the nuclear reaction network in the $\gamma$ process. We confirm that the current rate determination indicates that the possible uncertainty in the reaction cross section is small enough that it does not remarkably affect evolution of $^{161}$Ho or $^{160}$Dy abundances in the $\gamma$ process. Accordingly, the change of the reaction rate does not affect yields of nuclei with $A <160$ produced through the pathway of $^{161}$Ho($\gamma, n$)$^{160}$Dy.

\begin{figure}
\begin{center}
\includegraphics[width=0.48\textwidth]{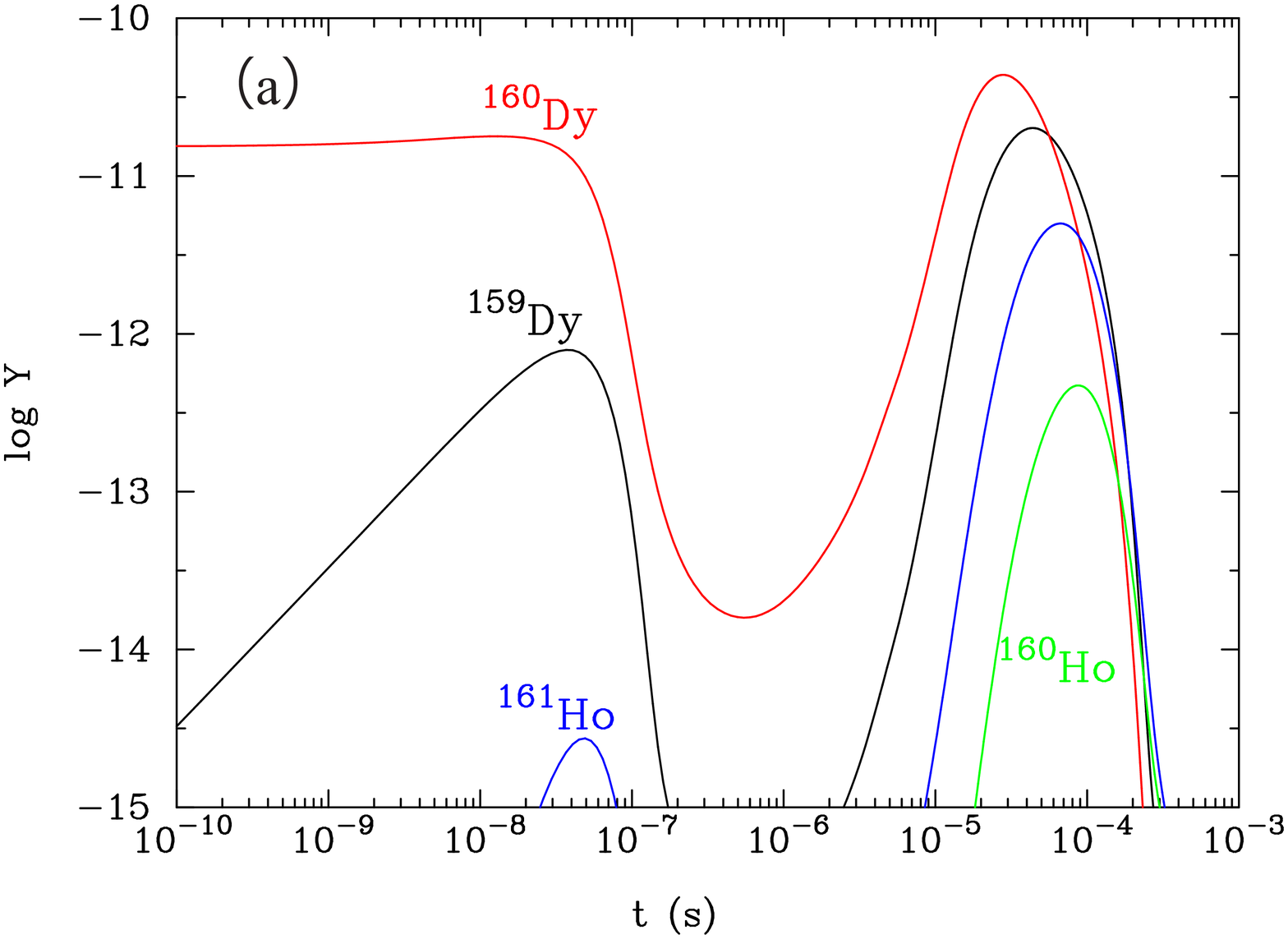}
\includegraphics[width=0.48\textwidth]{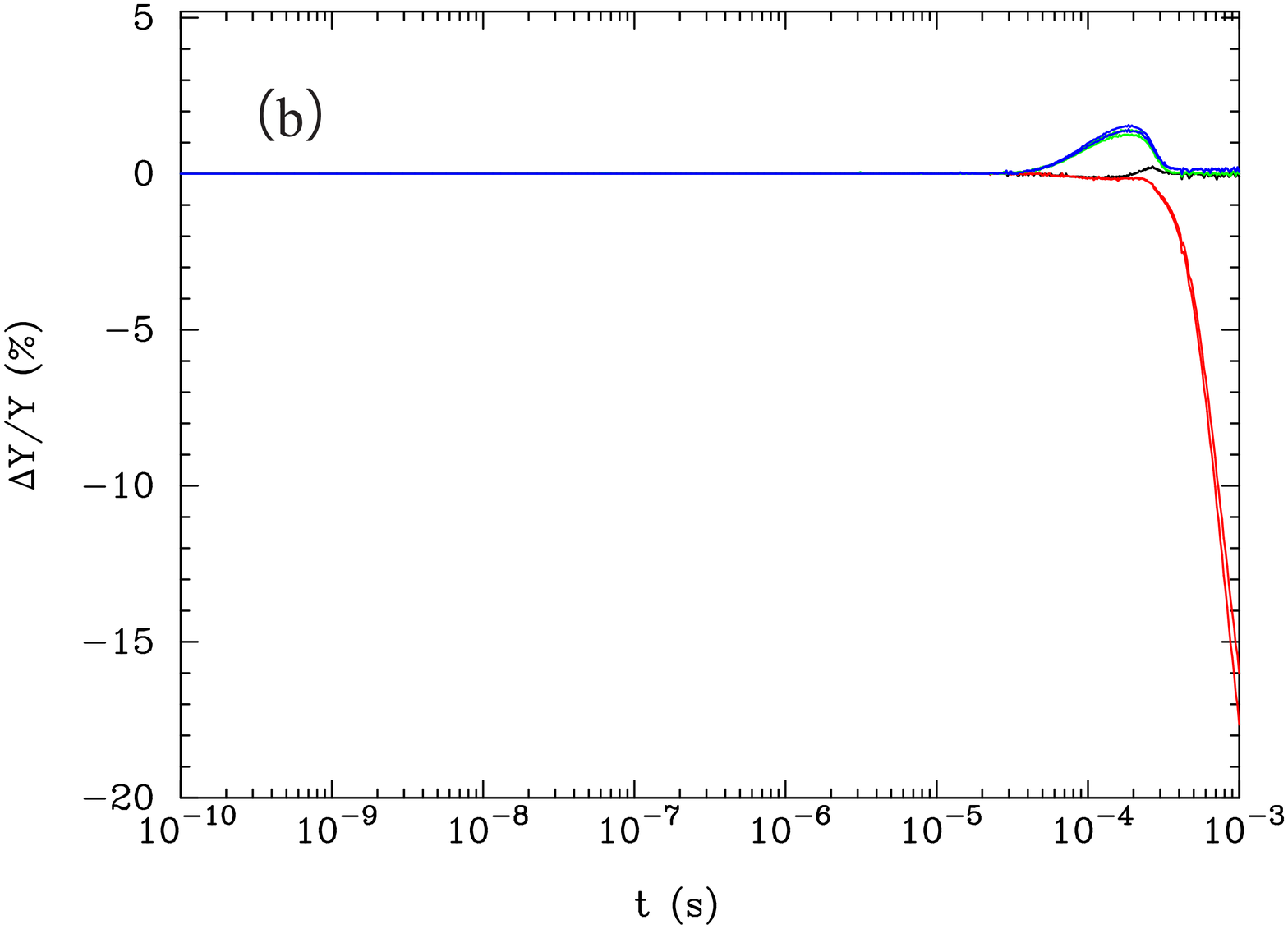}
\end{center}
\caption{(a) Time evolution of nuclear mole fractions on the test trajectory with $T_{9p}=3$. (b) Fractional differences in the mole fractions between the standard case and the case of JINA REACLIB rate for $^{161}$Ho($\gamma, n$)$^{160}$Dy. Each narrow band that is almost invisible corresponds to a variation caused by the uncertainty in the evaluated rate.
  \label{fig_ppro2}}
\end{figure}

Figure~\ref{fig_ppro2} (a) shows time evolution of abundances on the test trajectory with $T_{9p}=3$. The temperature is rather high, and first neutrons are released shortly. Instantaneous neutron captures make a flow of nuclear abundances to the neutron-rich region on the nuclear chart at $t \sim 10^{-7}$ s, and abundances of $^{159,160}$Dy and $^{161}$Ho decrease. After the neutron abundance decreases, successive ($\gamma,n$) reactions move back the flow toward the proton-rich region. Then, abundances of $^{159, 160}$Dy and $^{160, 161}$Ho  increase at $t \sim 10^{-5}$--$10^{-4}$ s and eventually decrease due to the photodisintegration into a more proton-rich and low $A$ region.
 Figure~\ref{fig_ppro2} (b) shows fractional differences in the abundances between the standard case and the case of the JINA REACLIB rate for $^{161}$Ho($\gamma, n$)$^{160}$Dy. Each band bounded by two lines delineates a variation caused by the uncertainty in the evaluated rate. Variations of ${\mathcal O}(1)$ \% are seen when the photodisintegration is effectively destroying those nuclei. The smaller $^{161}$Ho($\gamma, n$)$^{160}$Dy rate in our standard case at $T_9= 3$ increases abundances of $^{161}$Ho which survives that reaction and $^{160}$Ho produced via $^{161}$Ho($\gamma,n$)$^{160}$Ho. Also, the $^{160}$Dy abundance slightly decreases due to the hindered production via the $^{161}$Ho($\gamma, n$)$^{160}$Dy. The uncertainty in the current rate corresponds to at most ${\mathcal O}(0.1)$ \% differences in abundances.

\begin{figure}
\begin{center}
\includegraphics[width=0.48\textwidth]{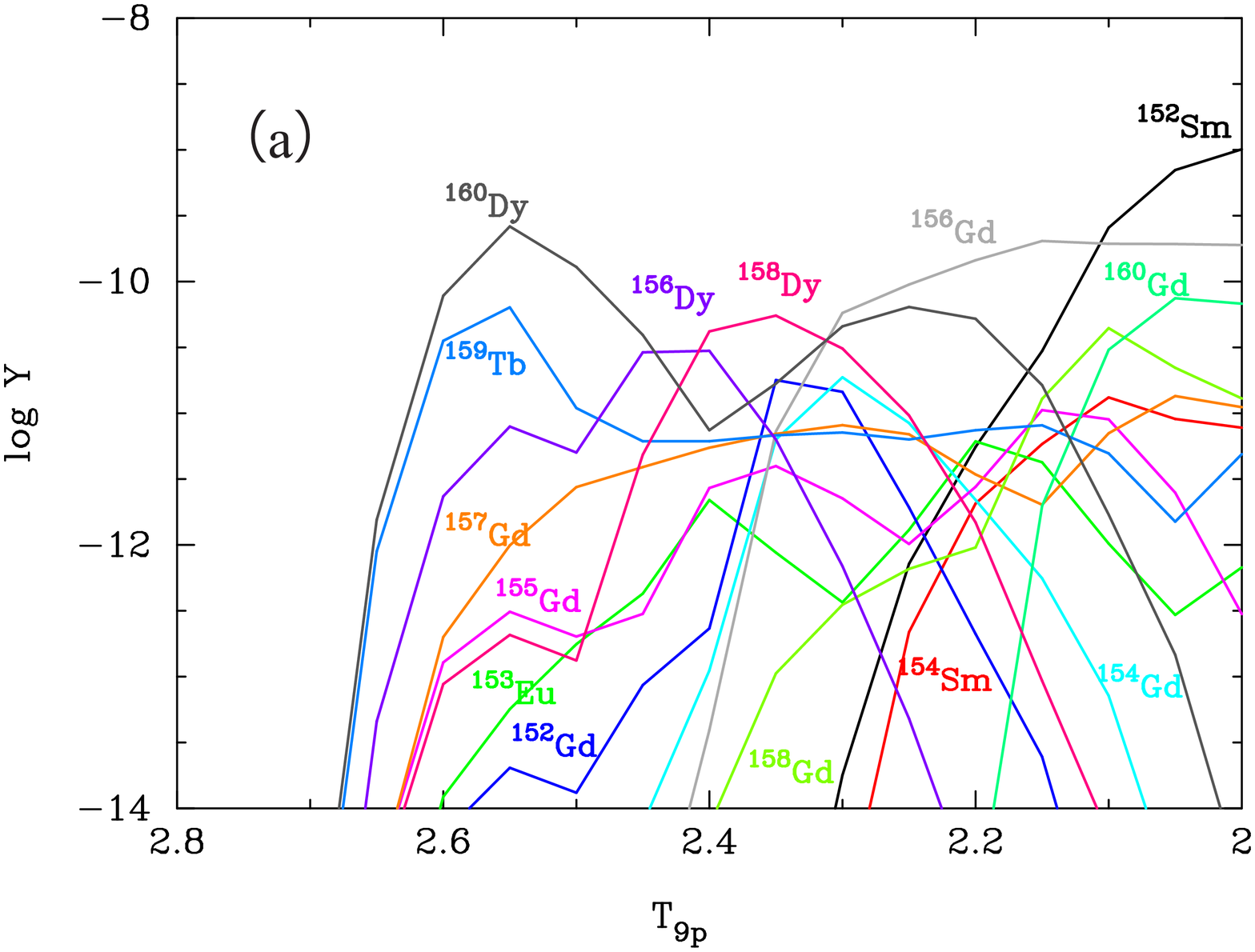}
\includegraphics[width=0.48\textwidth]{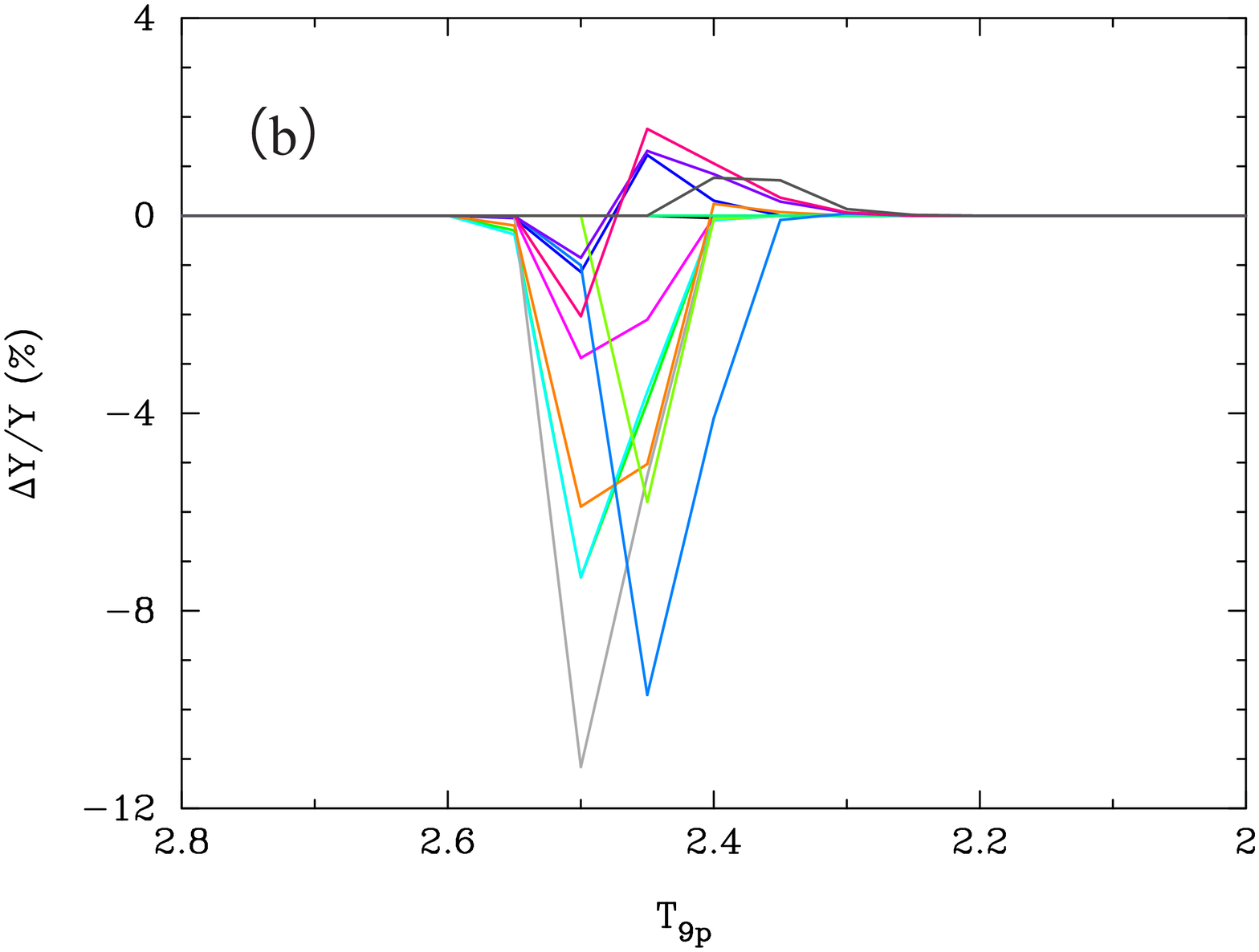}
\end{center}
\caption{(a) Mole fractions of stable nuclei with mass number $A \lesssim 160$ in the $\gamma$ process in SN Ia as a function of the peak temperature $T_{9p}$. (b) Fractional differences in the final mole fractions in the case with the $^{161}$Ho($\gamma, n$)$^{160}$Dy rate enhanced by a factor of 100 compared to the standard case.
  \label{fig_ppro1}}
\end{figure}


The effect of the change in the $^{161}$Ho($\gamma, n$)$^{160}$Dy rate is small even at $t \sim 10^{-5}$--$10^{-4}$ s when the photodisintegration operates predominantly. This is because the $^{161}$Ho($\gamma,n$) rate is much larger than the ($\gamma, n$) rate for $T_9 \geq 2$ relevant to the  $\gamma$ process. At such high temperatures of $T_9 \lesssim 3$, all heavy seed nuclei are disintegrated, and yields are zero (Figure \ref{fig_ppro1}). At lower temperatures, although the changes in abundances could remain in the final abundances, those are constrained to be at ${\mathcal O}(0.1)$ \% level.

Figure \ref{fig_ppro1} (a) shows final yields of stable nuclei with mass numbers $A \lesssim 160$ in the $\gamma$ process in SN Ia as a function of $T_{9p}$.
Since those nuclei are located immediately downstream of the pathway $^{161}$Ho($\gamma, n$)$^{160}$Dy, changes are expected in abundances of those nuclei if that reaction is important in determining the final abundances.
As a standard case, we adopt the rate for $^{161}$Ho($\gamma, n$)$^{160}$Dy derived from the $^{160}$Dy Best parameters. Figure \ref{fig_ppro1}  (b) shows fractional differences in the final yields between the case with the $^{161}$Ho($\gamma, n$)$^{160}$Dy rate 100 times as large as the adopted rate and the standard case. Even when the rate is thus extremely large, the yields vary by at most ${\mathcal O}(10)$ \%. Since the uncertainty in the $^{161}$Ho($\gamma, n$)$^{160}$Dy rate is estimated to be much smaller (Figure ~\ref{fig7}), it is found that the current determination is sufficiently precise to obtain accurate results of the  $\gamma$ process.

\section{Conclusion}

In this work we determined for the first time the cross sections of $^{160}$Dy($p,\gamma$)$^{161}$Ho and $^{161}$Dy($p,n$)$^{161}$Ho in the energy range of 3.4 to 7.0 MeV using the activation techniques.
This measurement extends the scarce experimental database for charged-particle-induced reactions on neutron-deficient nuclei.

The measured cross sections were compared to statistical model calculations obtained from the widely used Hauser-Feshbach codes: NON-SMOKER, EMPIRE, and TALYS.
Using the default input parameters it was found that cross sections for both reaction channels agreed with TALYS theoretical prediction.
The results allowed us to constrain strongly the $\gamma$ strength function and nuclear level density used in the HF statistical model.

With the TALYS code, we find the optimized TALYS input for the $A \sim $ 160 mass range, and further recommend the stellar rates for $^{161}$Ho($\gamma, n$)$^{160}$Dy over a large temperature range for $\gamma$ process network calculations. This approach is practically useful to compute the reaction rates in short-range extension to the experimentally known region.
We would like to point out that the large uncertainties in the weak $\gamma$ intensity of interest in Dy isotopes,
impedes the improvement of accuracy in cross-section determinations.
A dedicated experiment to further improve the gamma intensity will help to reduce the uncertainty.

Adopting the derived rate for the photodisintegration $^{161}$Ho($\gamma, n$)$^{160}$Dy, evolutions of nuclear abundances around $A=160$ in the  $\gamma$ process were analyzed. We conclude that employing the determined $^{161}$Ho($\gamma, n$)$^{160}$Dy rate
does not affect the yields of nuclei with $A \sim  160$. Moreover, a sensitivity study shows that the present experimental cross section is precise enough for the $\gamma$ process calculations.

The low yields make the proton or $\alpha$ irradiation measurements relevant for $\gamma$ process very time consuming. A possible way to improve this is to use the stacked target activation method (e.g., \citep[]{2020EPJA...56...13M} for measurements of cross sections at energies lower than the provided beam energy. If there are $\gamma$ activity setups available, simultaneous measurements of cross sections at a series of energies would be possible.

\begin{acknowledgments}
Many thanks to Dr. B. Mei for helpful discussions on the reaction rates.
This work was supported in part by the National Natural Science Foundation of China (Nos.11575018, U1867210, U1832211, 11961141004, 11922501, and 11790322) and by the National Key R \& D program of China (No. 2016YFA0400504).

\end{acknowledgments}


\appendix
 
\section{ The derivation of cross section}\label{crosssection}

The activation method is constituted by the irradiation and the residual measurements of the experimental target. The rate of change in the number of radioactive nuclei is given by the difference of production and decay rate,
\begin{equation}
\frac{dN(t)}{dt}=\sigma(E)N_s I(t)-\lambda N(t)\;,\qquad 0<t<t_b
 \end{equation}
 where $N(t)$ and $\lambda$ are the number and the decay constant of the object nucleus, $\sigma(E)$ the cross section of the reaction at the bombarding energy  $E$,  
 $N_s$ the number of target nuclei, and $I(t)$ is the beam intensity at a time $t$. $t_b$ is the irradiation time.

The number of reaction products is:
\begin{equation} 
N(t_b)=N_s\sigma(E)e^{-\lambda t_b}\int_0^{t_b}I(t)e^{\lambda t}dt\;,
\label{eq3}
\end{equation}

In the present work, a waiting time $t_w$ was needed to release the vacuum, dismount the target and place the target in the position for off-line measurement. Then the targets were measured for $t_m$ (measurement time). The decayed $\gamma$ rays emitted from the target is thus
\begin{equation}
n_\gamma=N(t_b)e^{-\lambda t_w}(1-e^{-\lambda t_m})\varepsilon_\gamma\eta_\gamma\;,
\label{eq4}
\end{equation}
 where $\varepsilon_\gamma$ and $\eta_\gamma$ are the detection efficiency and the gamma intensity, respectively.  

Deduced from Eq. \ref{eq3} and Eq. \ref{eq4}, the cross section of the reaction, $\sigma(E)$, is
\begin{equation} 
\sigma(E)=\frac{ n_\gamma}{N_se^{-\lambda t_b}e^{-\lambda t_w}(1-e^{-\lambda t_m})\varepsilon_\gamma \eta_\gamma\int_0^{t_b}I(t)e^{\lambda t}dt}\;,
\label{eq5}
\end{equation}

For the special case of a constant flux, $I(t)$ =  $I_0$, the above equations can be solved analytically. Eq.~\ref{eq5} can be rewritten to
\begin{equation}
\sigma(E)=\frac{\lambda n_\gamma}{N_sI_0e^{-\lambda t_w}(1-e^{-\lambda t_m}) (1-e^{-\lambda t_b})\varepsilon_\gamma \eta_\gamma}\;
\label{eq:sigmaE}
\end{equation}
 
\end{document}